\documentclass{article}
\usepackage{spconf,amsmath,graphicx}

\usepackage{lipsum} 

\usepackage{musicography}
\usepackage{enumitem}
\usepackage{siunitx}
\usepackage{hyperref}
\usepackage{cleveref}
\usepackage{footmisc}
\usepackage{booktabs}
\graphicspath{{images/}}

\usepackage{xcolor}
\definecolor{rowblue}{RGB}{220,230,240}
\definecolor{myorchid}{RGB}{150,10,30}
\definecolor{myblue}{RGB}{10,30,250}
\definecolor{mygreen}{RGB}{10,190,10}
\definecolor{myred}{RGB}{190,20,20}

\title{Differentiable Signal Processing With Black-Box Audio Effects}
\name{Marco A. Mart\'{i}nez Ram\'{i}rez$^{\musFlat{}\musSharp{}}$\sthanks{This work was performed while interning at Adobe Research.} \quad Oliver Wang$^{\musSharp{}}$ \quad Paris Smaragdis$^{\musNatural{}\musSharp{}}$ \quad Nicholas J. Bryan$^{\musSharp{}}$}
\address{$^{\musSharp{}}$Adobe Research, USA\\
$^{\musFlat{}}$Centre for Digital Music, Queen Mary University of London, UK\\
 $^{\musNatural{}}$University of Illinois at Urbana-Champaign, USA}
\begin{document}
\ninept
\maketitle

\begin{abstract}
We present a data-driven approach to automate audio signal processing by incorporating stateful third-party, audio effects as layers within a deep neural network. 
We then train a deep encoder to analyze input audio and control effect parameters to perform the desired signal manipulation, requiring only input-target paired audio data as supervision. 
To train our network with non-differentiable black-box effects layers, we use a fast, parallel stochastic gradient approximation scheme within a standard auto differentiation graph, yielding efficient end-to-end backpropagation. 
We demonstrate the power of our approach with three separate automatic audio production applications: tube amplifier emulation, automatic removal of breaths and pops from voice recordings, and automatic music mastering. 
We validate our results with a subjective listening test, showing our approach not only can enable new automatic audio effects tasks, but can yield results comparable to a specialized, state-of-the-art commercial solution for music mastering.
\end{abstract}


\section{Introduction}
\label{sec:intro}
Audio signal processing effects (Fx) are ubiquitous and used to manipulate different sound characteristics such as loudness, dynamics, frequency, and timbre across a variety of media. 
Many Fx, however, can be difficult to use or are simply not powerful enough to achieve a desired task, motivating new work. 
Past methods to address this include audio effects circuit modeling~\cite{yeh2008numerical}, analytical methods~\cite{eichas2016black}, and intelligent audio effects that dynamically change their parameter settings by exploiting sound engineering best practices~\cite{IMPbook19}. 
The most common approach for the latter is adaptive audio effects or signal processing systems based on the modeling and automation of traditional processors~\cite{moffat2019approaches}. 
More recent deep learning methods for audio effects modeling and intelligent audio effects include 1) end-to-end direct transformation methods~\cite{martinez2020deep,wright2020real,hawley2019signaltrain}, where a neural proxy learns and applies the transformation of an audio effect target 2) parameter estimators, where a deep neural network (DNN) predicts the parameter settings of an audio effect \cite{ramo2019neural,sheng2019feature, mimilakis2020one} and 3) differentiable digital signal processing (DDSP)~\cite{engel2020ddsp,kuznetsov2020differentiable}, where signal processing structures are implemented within a deep learning auto-differentiation framework and trained via backpropagation. 

While these past methods are very promising, they are limited in several ways. 
First, direct transform approaches can require special, custom modeling strategies per effect (e.g. distortion), are often based on large and expensive networks, and/or use models with limited or no editable parameter control~\cite{martinez2020deep, wright2020real}.
Second, methods with parameter control commonly require expensive human-labeled data from experts (e.g. inputs, control parameters, and target outputs)~\cite{hawley2019signaltrain}. They are also typically optimized to minimize parameter prediction error and not audio quality directly~\cite{mimilakis2020one}, which can reduce performance. 
Third, DDSP approaches require a differentiable implementation for learning with backpropagation, re-implementation of each Fx, and in-depth knowledge, limiting use to known differentiable effects and causing high engineering effort~\cite{engel2020ddsp,kuznetsov2020differentiable}.  

\begin{figure}[t!]
\centering
\makebox[0pt]{\includegraphics[trim=.35cm 0.95cm 0 0.5cm, width=1\linewidth]{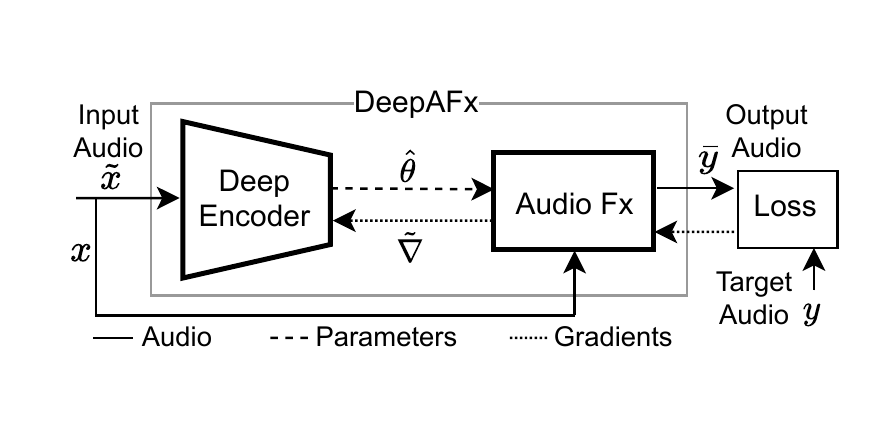}}
\caption{\label{fig:model}{Our \textit{DeepAFx} method consists of a deep encoder that analyzes audio and predicts the parameters of one or more black-box audio effects (Fx) to achieve a desired audio processing task.
At training time, gradients for black-box audio effects are approximated via a stochastic gradient method.
}}
\end{figure}

In this work, we introduce \textit{DeepAFx}, a new approach to differentiable signal processing that enables the use of \emph{arbitrary} stateful, third-party black-box processing effects (plugins) within any deep neural network as shown in~\Cref{fig:model}. 
Using our approach, we can mix-and-match traditional deep learning operators and black-box effects and learn to perform a variety of intelligent audio effects tasks without the need of neural proxies, re-implementation of Fx plugins, or expensive parameter labels (only paired input-target audio data is required).
To do so, we present 1) a deep learning architecture in which an encoder analyzes input audio and learns to control Fx black-boxes that themselves perform signal manipulation 2) an end-to-end backpropagation method that allows differentiation through non-differentiable black-box audio effects layers via a fast, parallel stochastic gradient approximation scheme used within a standard auto differentiation graph 3) a training scheme that can support stateful black-box processors, 4) a delay-invariant loss to make our model more robust to effects layer group delay, 5) application of our approach to three audio production tasks including tube amplifier emulation, automatic non-speech sounds removal, and automatic music post-production or mastering, and 6) listening test results that shows our approach can not only enable new automatic audio production tasks, but is high-quality and comparable to a state-of-the-art proprietary commercial solution for automatic music mastering. Audio samples and code are online at \url{https://mchijmma.github.io/DeepAFx/}.

\section{Method}
\label{sec:methods}

\subsection{Architecture}
\label{sec:model}
Our model is a sequence of two parts: a deep encoder and a black-box layer consisting of one or more audio effects.
The encoder analyzes the input audio and predicts a set of parameters, which is fed into the audio effect(s) along with the input audio for transformation. 

The \textbf{deep encoder} can be any deep learning architecture for learning representations of audio.
To allow the deep encoder to learn long temporal dependencies, the input $\tilde{x}$ consists of the current audio frame $x$ centered within a larger audio frame containing previous and subsequent context samples. 
The input to the encoder consists of a log-scaled mel-spectrogram non-trainable layer followed by a batch normalization layer.
The last layer of the encoder is a dense layer with $P$ units and \textit{sigmoid} activation, where $P$ is the total number of parameters. 
The output is the predicted parameters $\hat{\theta}$ for the current input frame $x$. 

The \textbf{audio Fx} layer is a stateful black-box consisting on one or more connected audio effects.
This layer uses the input audio $x$ and $\hat{\theta}$ to produce an output waveform $\bar{y} = f(x, \hat{\theta})$. We calculate a loss between the transformed audio and targeted output audio.

\subsection{Gradient approximation}
\label{sec:gradient}

In order to train our model via backpropagation, we calculate the gradients of the loss function with respect to the audio Fx layer and the deep encoder.
For the latter, we compute the gradient using standard automatic differentiation. 
For the former, gradient approximation has been addressed in various ways such as finite differences (FD)~\cite{milne2000calculus}, FD together with evolution strategies~\cite{salimans2017evolution}, 
reinforcement learning~\cite{williams1992simple}, and by employing neural proxies~\cite{jacovi2019neural}.  
For our work, we use a stochastic gradient approximation method called simultaneous permutation stochastic approximation (SPSA)~\cite{spall1998overview}, which is rooted in standard FD gradient methods and has low computational cost~\cite{milne2000calculus}.
In our formulation, we ignore the gradient of the input signal $x$ and only approximate the gradient of the parameters $\hat{\theta}$.

\textbf{FD} can be used to approximate the gradient of the audio Fx layer at a current set of parameters $\hat{\theta}_{0}$, where the gradient $\tilde{\nabla}$ is defined by the vector of partial derivatives characterized by $\tilde{\nabla} f(\hat{\theta}_{0})_{i} = \partial f(\hat{\theta}_{0})/\partial\theta_{i}$ for $i = 1,\dots,P$. 
The two-side FD approximation $\tilde{\nabla}^{FD}$ \cite{spall1998overview} is based on the backward and forward measurements of $f()$ perturbed by a constant $\epsilon$. 
The $i$th component of $\tilde{\nabla}^{FD}$ is
\begin{equation}
\tilde{\nabla}^{FD}f(\hat{\theta}_{0})_{i} = \frac{f(\hat{\theta}_{0}+\epsilon \hat{d}^{P}_{i})-f(\hat{\theta}_{0}-\epsilon \hat{d}^{P}_{i})}{2\epsilon},
\label{eq:FD}
\end{equation}
where $0< \epsilon \ll 1$ and $\hat{d}^{P}_{i}$ denotes a standard basis vector with $1$ in the $i$th place and dimension $P$. From \cref{eq:FD}, each parameter $\theta_{i}$ is perturbed one at a time and requires  2$P$ function evaluations. 
This can be computationally expensive for even small $P$, given that $f()$ represents one or more audio effects. Even worse, when $f()$ is stateful, FD perturbation requires $2P$ instantiations of $f()$ to avoid state corruption, resulting in high memory costs.

\textbf{SPSA} offers an alternative and is an algorithm for stochastic optimization of multivariate systems. It has proven to be an efficient gradient approximation method that can be used with gradient descent for optimization~\cite{spall1998overview}. The SPSA gradient estimator $\tilde{\nabla}^{SPSA}$ is based on the random perturbation of all the parameters $\hat{\theta}$ at the same time. The $i$th element of $\tilde{\nabla}^{SPSA}$ is
\begin{equation}
\tilde{\nabla}^{SPSA}f(\hat{\theta}_{0})_{i} = \frac{f(\hat{\theta}_{0}+\epsilon \hat{\Delta}^{P})-f(\hat{\theta}_{0}-\epsilon \hat{\Delta}^{P})}{2\epsilon\Delta^{P}_{i}},
\label{eq:SPSA}
\end{equation}
where $\hat{\Delta}_{P}$ is a $P$-dimensional random perturbation vector sampled from a symmetric Bernoulli distribution, i.e. $\Delta^{P}_{i}=\pm1$ with probability of 0.5 \cite{spall1992multivariate}. 
In each iteration, the total number of function evaluations $f()$ is two, since the numerator in \cref{eq:SPSA} is identical for all the elements of $\tilde{\nabla}^{SPSA}$. 
This is especially valuable when $P$ is large and/or when $f()$ is stateful.
While the random search directions do not follow the steepest descent, over many iterations, errors tend to average to the sample mean of the stochastic process \cite{spall1998overview}. 

\subsection{Training with stateful black-boxes}
\label{ssec:sequential}

While training samples typically should be generated from an independent and identically distributed (i.i.d.) process, we note that audio effects are \emph{stateful} systems. This means output samples depend on previous input samples or internal states~\cite{zolzer2011dafx}.
Therefore, audio Fx layers should not be fed i.i.d.-sampled audio during training as this will corrupt state and yield unrealistic, transient outputs. As such, we feed consecutive non-overlapping frames of size $N$ to the audio Fx layer, i.e. audio frames with a hop size of $N$ samples, where the internal block size of each audio effects is set to a divisor of $N$. This ensures that there are no discrepancies between the behavior of the audio effects during training and inference time.

To train the model using mini-batches and maintain state, we use a separate Fx processor for each item in the batch. 
Therefore, for a batch size of $M$, we instantiate $M$ independent audio effects for the forward pass of backpropogation.
Each of the $M$ plugins receives training samples from random audio clips, however across minibatches, each instance is fed samples from the same input until swapped with a new sample. 
When we consider the two-side function evaluations for SPSA, we meet the same stateful constraints by using two additional Fx per batch item. This gives a total of 3$M$ audio effects when optimizing with SPSA gradients, whereas FD requires an unmanageable $(\textrm{2}P + \textrm{1})M$ effects for a large number of parameters $P$ or batch size $M$. 
Finally, we parallelize our gradient operator per item in a minibatch, similar to recent distributed training strategies~\cite{you2018imagenet}, but using a single one-GPU, multi-CPU setup.

\subsection{Delay-invariant loss}
\label{ssec:loss}

Multiband audio effects correspond to audio processors that split the input signal into various frequency bands via different types of filters~\cite{zolzer2011dafx}.
These filters can introduce group delay, which is a frequency-dependent time delay of the sinusoidal components of the input~\cite{smith2007introduction}. 
These types of effects and similar can also apply a $\ang{180}$ phase shift, i.e. to invert the sign of the input. While such properties are critical to the functioning of Fx, it means that difficulties can arise when directly applying a loss function with inexact inputs, either in time or frequency domain. 
Thus, we propose a delay-invariant loss function designed to mitigate these issues.

We first compute the time delay $\tau = \operatorname*{argmax} (\bar{y} \star y)$ between the output $\bar{y}$ and target $y$ audio frames via cross-correlation ($\star$). 
Then, the loss in the time domain 
\begin{equation}
L_{time} = \operatorname{min} (||\bar{y}_{\tau}-y_{\tau}||_{1},||\bar{y}_{\tau}+y_{\tau}||_{1}),
\label{eq:time_domain_loss}
\end{equation}
corresponds to the minimum $L1$ distance between the time-aligned target ${y}_{\tau}$ and both a $\ang{0}$ phase shift and $\ang{180}$ phase shift time-aligned output $\bar{y}_{\tau}$.
We compute $\bar{Y}_{\tau}$ and $Y_{\tau}$: a 1024-point Fast Fourier Transform (FFT) magnitude of $\bar{y}_{\tau}$ and $y_{\tau}$, respectively. 
The loss in the frequency domain $L_{freq}$ is defined as
\begin{equation}
L_{freq} = ||\bar{Y}_{\tau}-Y_{\tau}||_{2} + ||\log\bar{Y}_{\tau}-\log Y_{\tau}||_{2}.
\label{eq:lossa}
\end{equation}
The final loss function is $L = \alpha_1 L_{time} + \alpha_2 L_{freq}$, where we empirically tune the weighting and use $\alpha_1=10$ and $\alpha_2=1$.

\begin{figure*}[t]
\centering
\begin{tabular}{*{10}{c@{\hspace{1.0pt}}}}
{\includegraphics[trim={3.cm 2.1cm 0 4.cm},clip,height=4.4cm,width=0.33\textwidth]{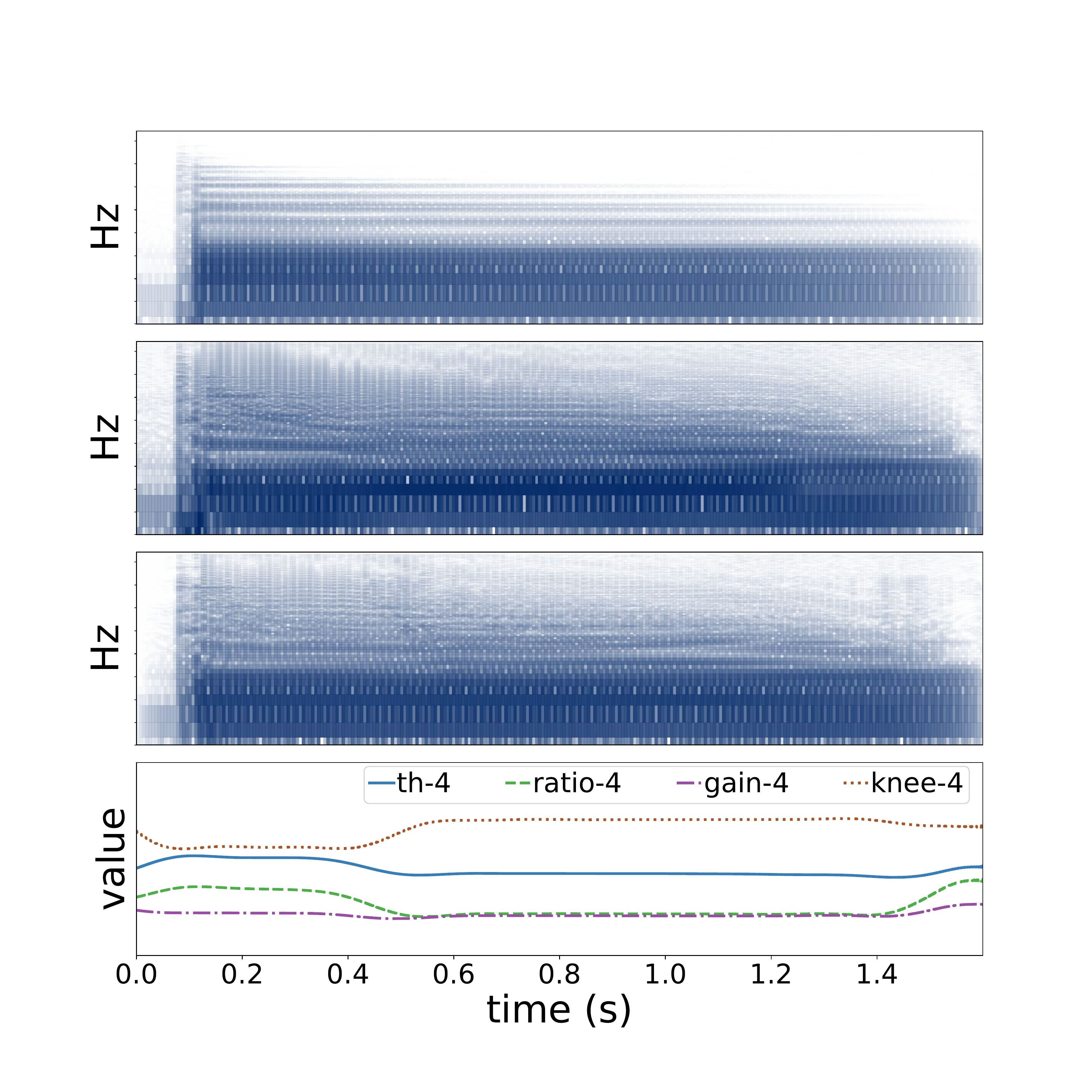}\label{fig:speech_spec}} &
{\includegraphics[trim={3.cm 2.1cm 0 4.cm},clip,height=4.4cm,width=0.33\textwidth]{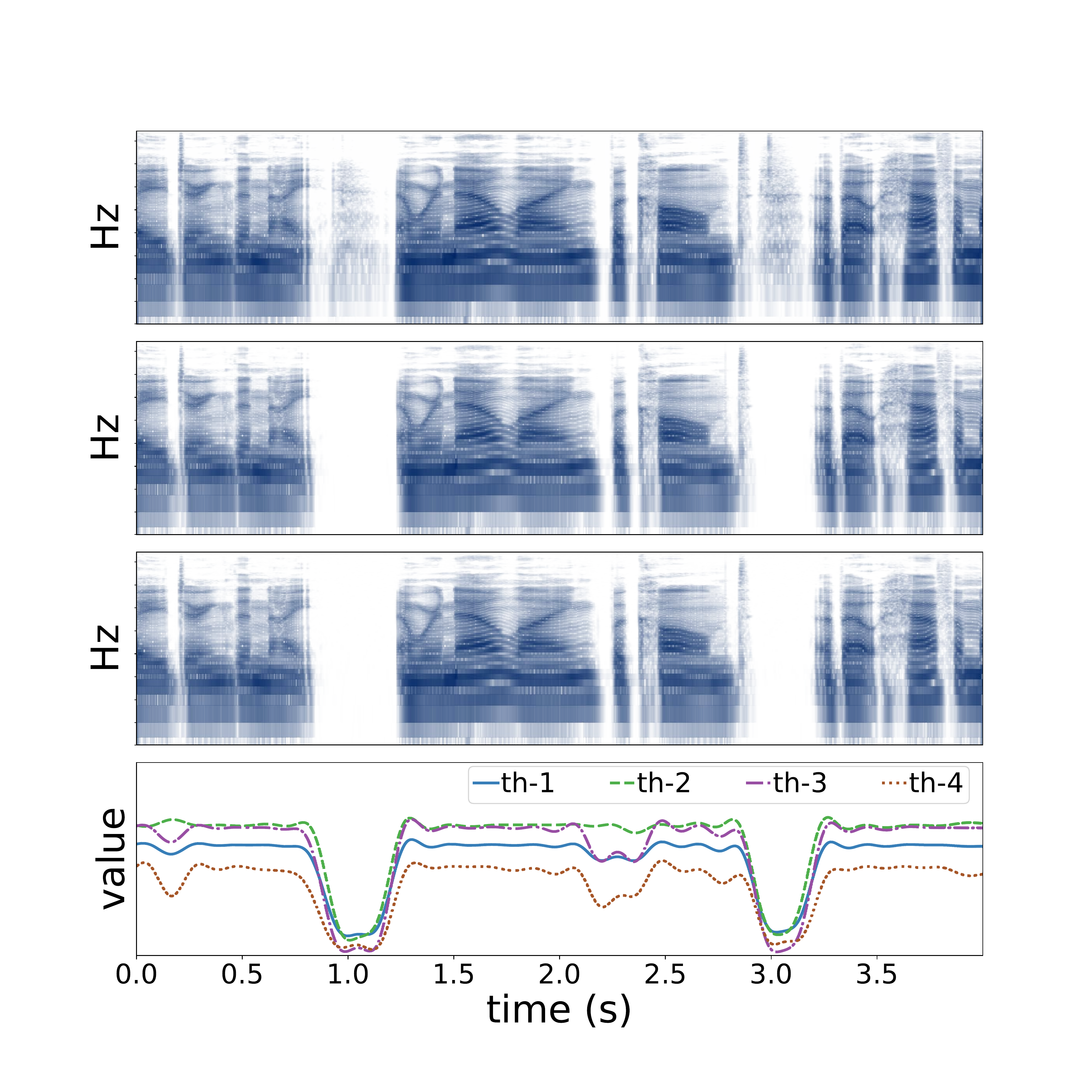}\label{fig:distortion_spec}} &
{\includegraphics[trim={3.cm 2.1cm 0 4.cm},clip,height=4.4cm,width=0.33\textwidth]{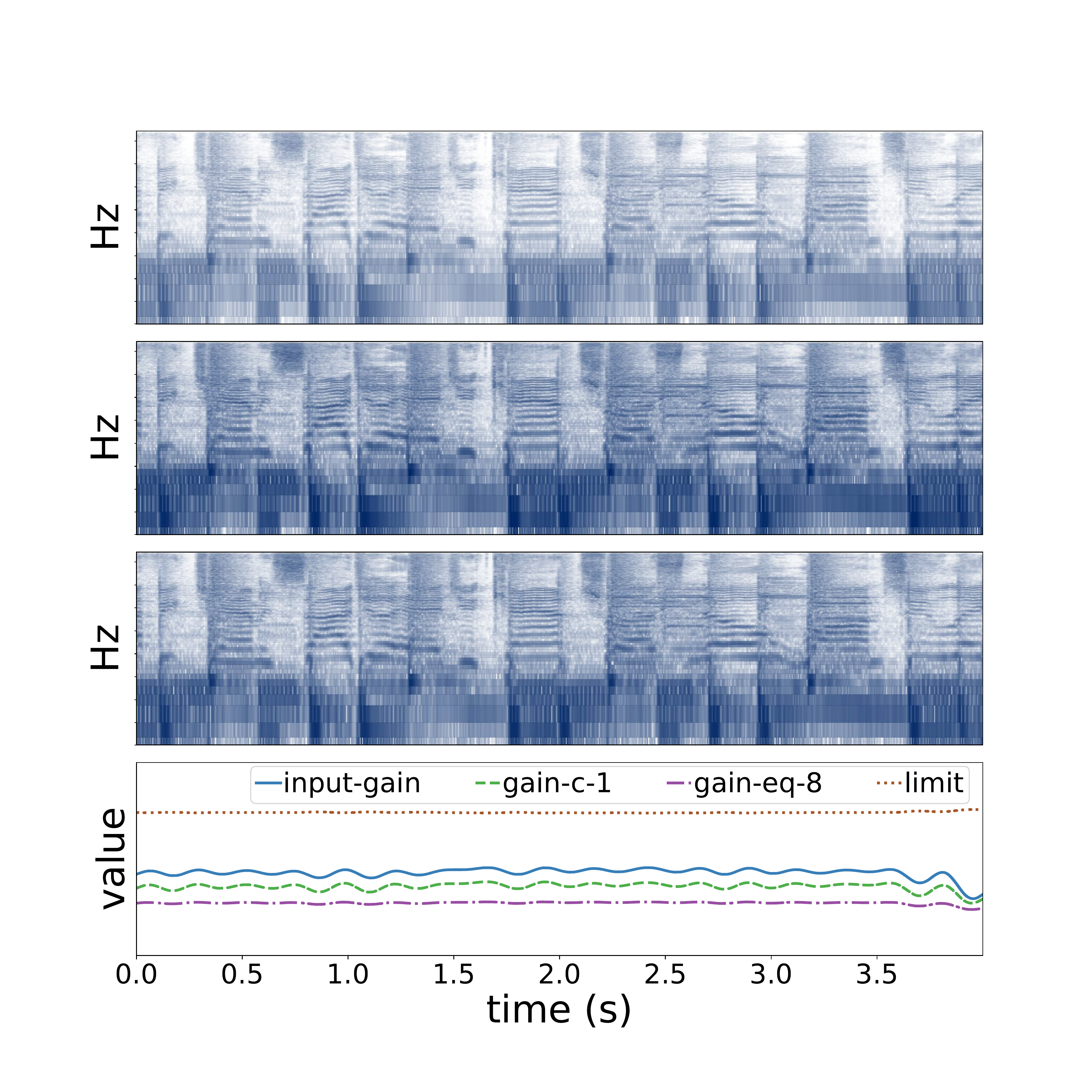}\label{fig:mastering_spec}} \\
\end{tabular}
\vspace{-.4cm}
\caption{(Left) Tube amplifier emulation. (Center) Automatic non-speech sounds removal. (Right) Automatic music mastering. From top to bottom: input, target, our result, and selected predicted parameters. \textit{th}, and \textit{gain-c} mean threshold and compressor makeup gain, respectively.}
\label{fig:frames}
\end{figure*}

\section{Experiments}
\label{sec:experiments}

We apply our approach to three applications including tube amplifier emulation, automatic non-speech sounds removal, and automatic music mastering. In each case, we use one or more different effects and different datasets to illustrate the generality of our approach.
We also explore two deep encoder variants; an \textit{Inception} network~\cite{lee2020metric} and \textit{MobileNetV2}~\cite{sandler2018mobilenetv2}. 
The number of parameters for each encoder is approx. $2.8$M and $2.2$M, respectively. The input context $\tilde{x}$ and current audio frame $x$ are 40960 and 1024 samples (1.85 sec and 46 ms) at a 22,050Hz sampling rate. 
The log-scaled mel-spectrogram input layer has a 46 ms window size, 25\% overlap, and 128 mel-bands. We use effects from the \textit{LV2} audio plugin open standard~\cite{lv2ref} and experiment with continuous parameters scaled between 0 and 1. At inference we apply a lowpass filter to $\hat{\theta}$ to ensure smoothness. For intuition, our final optimization scheme with the Inception encoder and parametric equalizer Fx (EQ)~\cite{lsppeqref} takes 3 minutes per epoch (1000 steps) to train with batch size $M=100$ on a Tesla-V100 GPU. 
We use Adam optimization and early stop on validation loss.

\subsection{Tube amplifier emulation}
\label{ssec:distortion}

The emulation of distortion effects is a highly-researched field~\cite{yeh2008numerical,eichas2016black} whose goal is to recreate the sound of analog reference devices (e.g. tube amplifiers or distortion circuits). 
Amplifier emulation using deep learning has been explored in the context of deep neural proxies~\cite{martinez2019modeling,damskagg2019deep} and been shown to achieve virtually indistinguishable emulation under certain conditions~\cite{martinez2020deep,wright2020real}. 
Such approaches, however, are \emph{specifically} designed for this task, making it an interesting benchmark. We ask -- is it possible to emulate distortion using a simple multiband dynamic range compressor? 

A compressor modifies the amplitude dynamics of audio by applying a time-varying gain. 
Compressors are commonly used for loudness control and typically introduce little harmonic distortion in contrast to tube amplifiers~\cite{zolzer2011dafx}. 
Nonetheless, we apply our approach and learn 21 parameters; the \textit{threshold}, \textit{makeup gain}, \textit{ratio}, and \textit{knee} for each of the 4 frequency bands; the 3 \textit{frequency splits}; and the \textit{input} and \textit{output gains} on a multiband compressor~\cite{calfref}. 
For training, we use the subset of the IDMT-SMT-Audio-Effects dataset~\cite{stein2010automatic} used by Mart\'{i}nez et al.~\cite{martinez2020deep} consisting of 1250 raw notes from various electric guitars and bass guitars and processed through a \textit{Universal Audio 6176 Vintage Channel Strip} tube preamplifier. 
The train, validation, and test dataset sizes are 31.6, 4.0, and 4.0 minutes, respectively. 
For all tasks, the non-trainable Fx parameters are set to defaults with the exception of the attack and release gates, which are set to their minimum (10 ms) due to the fast control rate of our model (46 ms). 

\subsection{Automatic non-speech sound removal}
\label{ssec:speech}
Removing non-speech vocal sounds, such as breaths and lip smacks, is a common task performed by sound engineers~\cite{owsinski2013mixing}. 
This task can be done by manually editing the audio waveform or by using a \textit{noise gate}, to reduce signal below a certain \textit{threshold} via a \textit{reduction gain} and \textit{ratio} setting~\cite{zolzer2011dafx}. 
Both approaches can be time consuming, require expert knowledge, and, to the best of our knowledge, the automation of this task has not been investigated with the tangential exception of an intelligent noise gate for drum recordings~\cite{terrell2011automatic} and automatic detection of breaths for speaker recognition~\cite{dumpala2017algorithm, szekely2019casting}.

We train a model with a \textit{multiband noise gate}~\cite{calfref} Fx layer with 17 parameters (\textit{threshold}, \textit{reduction gain} and \textit{ratio} for each of the 4 frequency bands; the 3 \textit{frequency splits}; and the \textit{input} and \textit{output gains}). 
We use the \textit{DAPS} dataset~\cite{mysore2014can} which contains 100 raw and clean speech recordings with manually removed breaths and lip smacks. 
The train, validation, and test dataset size is 213.5, 30.2, and 23.8 minutes, respectively.

\subsection{Automatic music mastering}
\label{ssec:mastering}

Music post-production or mastering is the process of enhancing a recording by manipulating its dynamics and frequency content~\cite{IMPbook19}. 
This manipulation is typically done by an experienced mastering engineer and is carried out using dynamic range effects, such as a compressor and limiter; and frequency-based processors, such as EQ. Automatic mastering has been explored using adaptive audio effects \cite{mimilakis2013automated,hilsamer2014statistical}, using DNNs to predict dynamic range compression gain coefficients~\cite{mimilakis2016deep}, and even commercialized as proprietary online mastering software (OMS)~\cite{landrref}.

We train a model with multiple audio effects in series, similar to the way mastering engineers perform this task, consisting of a \textit{multiband compressor}~\cite{calfref}, a \textit{graphic EQ}~\cite{lsppeqref} and mono \textit{limiter}~\cite{lspplimitedref}. 
A limiter is a dynamic range processor that more aggressively adjusts any input above the threshold, e.g. by using a ratio of $\infty$~\cite{reiss2014audio}. 
We train our model to learn 50 parameters -- 16 parameters for the multiband compressor (\textit{threshold}, \textit{makeup gain} and \textit{ratio} for each of the 4 frequency bands, the 3 \textit{frequency splits}, and the \textit{input gain}), 33 parameters for the \textit{graphic EQ} (the \textit{gain} for each of the 32 frequency bands and the \textit{output gain}), and 1 parameter for the limiter (\textit{threshold}).
For training data, we collected 138 unmastered and mastered music tracks from \textit{The `Mixing Secrets' Free Multitrack Download Library}~\cite{senior2011mixing}, and as preprocessing step, we perform time-alignment using cross-correlation and loudness normalize each unmastered track to -25dBFS.
The train, validation, and test dataset size is 429.3,	51.1, and 50.3 minutes, and respectively.

\section{Results \& Analysis}
\label{sec:results}

\subsection{Qualitative Evaluation}
We show qualitative analysis by visual inspection via~\Cref{fig:frames}, in which we depict input, target and predicted spectrograms for all tasks. 
 For the tube amplifier task, it can be seen that the model emulates very closely the analog target. Also, the multiband compressor parameters move over time along the attack and decay of the guitar note to emulate the nonlinear hysteresis behavior of the tube amplifier. For the non-speech sound removal task, the model successfully removes breath and lip smacking sounds. It can be seen how the different thresholds vary according to the non-speech sounds present at the first and third second of the input recording. The music mastering example also successfully matches the sound engineer mastering target and the parameters of the three audio effects gradually change based on the input content of the unmastered music track. 

\begin{table}[t]
\centering
\resizebox{0.975\columnwidth}{!}{
\begin{tabular}{cc|ccc}
\toprule
	&	Model	&	Epochs	&	Time &	$\tilde{d}_{MFCC}$	\\
\midrule\midrule			
  Tube amplifier    &	Inception	&	97	&	9.07	&	0.2596	\\
emulation           &	MobileNetV2	&	63	&	6.4	&	0.2186	\\
        	        &	CAFx	&	723	&	5.5	&	0.0826	\\
\hline
Non-speech      &	Inception	&	89	&	7.4	&	0.0186	\\
sounds removal	&	MobileNetV2	&	60	&	4.8	&	0.0231	\\
\hline
Music 	    &	Inception	&	202	&	19.8	&	0.0282	\\
mastering	&	MobileNetV2	&	178	&	17.5	&	0.0542	\\
        	&	OMS	&	-	&	-	&	0.0157	\\
\bottomrule
\end{tabular}
}
\vspace*{-5pt}
\caption{Training epochs \& time (hours), and test MFCC distance. 
}
\vspace*{-5pt}
\label{table:test_loss}
\end{table}

\subsection{Quantitative Evaluation}
We show quantitative evaluation in \Cref{table:test_loss}, including number of training epochs until early stopping, training time in hours, and the mean cosine distance of the mel-frequency cepstral coefficients or $\tilde{d}_{MFCC}$ as a proxy for a perceptual metric, following past work~\cite{martinez2020deep}. 
For this, we compute 13 MFCCs from a log-power mel-spectrogram using a window size of 1024 samples, 25\% hop size and 128 bands.  As a baseline for the tube amplifier emulation task, we use the \textit{Convolutional Audio Effects Modeling Network (CAFx)} \cite{martinez2020deep}, and for the music mastering task, we use an online mastering software (OMS)~\cite{landrref}. We see that the tube amplifier emulation distance for our approach is higher than the other two tasks, likely caused by using a compressor to achieve distortion. We see both encoders achieved similar performance, although the Inception model tends to perform slightly better, and all training times are under a day. We also find the CAFx and OMS models have lower distance, but yield to perceptual listening tests for further conclusions. 

\subsection{Perceptual Evaluation}
We performed a subjective listening test to evaluate the perceptual quality of our results using the multiple stimulus hidden reference (MUSHRA) protocol~\cite{itu2014recommendation}. Seventeen participants took part in the experiment which was conducted using The Web Audio Evaluation Tool~\cite{jillings2015web}. Five participants identified as female and twelve participants identified as male.  Participants were among musicians, sound engineers, or experienced in critical listening.

\begin{figure}[t]
\begin{minipage}[b]{.32\columnwidth}
  \centering
  \centerline{\includegraphics[trim=0 0 0 0.25cm, clip, width=1.\columnwidth,height=2.1cm]{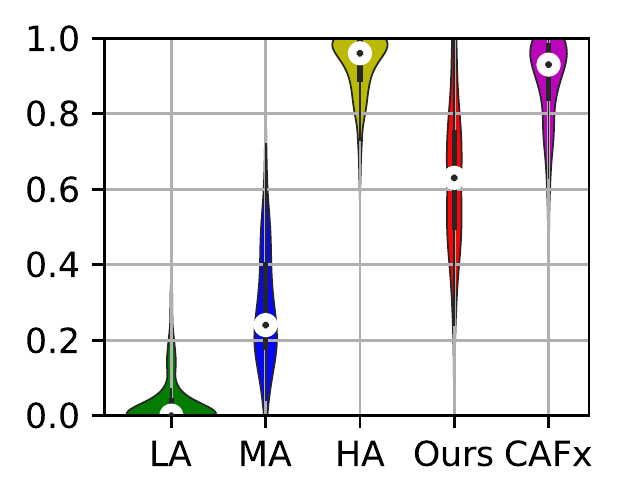}\label{fig:speech}}
\end{minipage}
\begin{minipage}[b]{.32\columnwidth}
  \centering
  \centerline{\includegraphics[trim=0 0 0 0.25cm, clip, width=1.0\columnwidth,height=2.1cm]{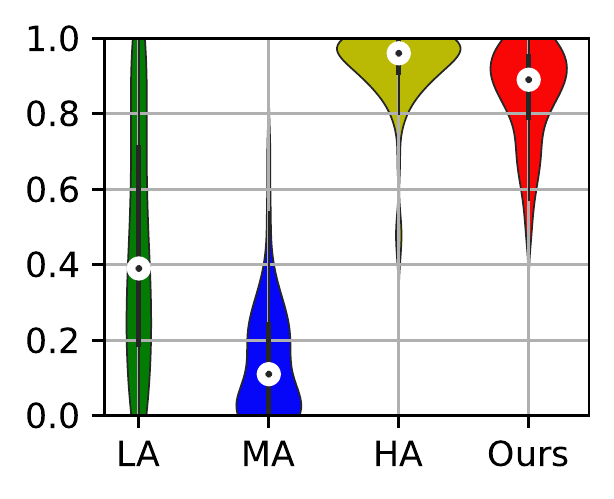}\label{fig:distortion}}
\end{minipage}
\begin{minipage}[b]{.32\columnwidth}
  \centering
  \centerline{\includegraphics[trim=0 0 0 0.25cm, clip, width=1\columnwidth,height=2.1cm]{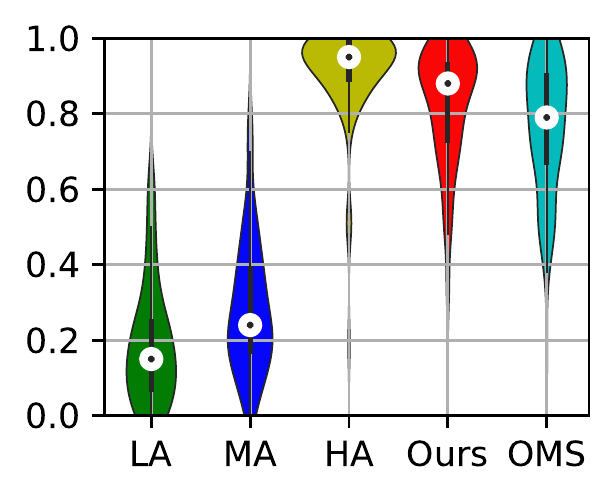}\label{fig:mastering}}
\end{minipage}
\vspace{-.35cm}
\caption{\label{fig:boxplot}{Listening test violins plots. (Left) Tube amplifier emulation results show our approach does not outperform CAFx~\cite{martinez2020deep}, but still yields good quality. (Center) Non-speech sounds removal results show near human quality. (Right) Music mastering results show our method performs equivalent to the state-of-the-art (OMS).}}
\end{figure}

Five listening samples were used from each of the three test subsets.  Each sample consisted of a 4-second segment with the exception of the tube amplifier emulation samples, where each note is 2 seconds long. Participants were asked to rate samples according to similarity relative to a reference sound, i.e. a sample processed by a sound engineer or a recording from an analog device. 

The samples consisted of an unprocessed sample as low-anchor (LA), a mid-anchor (MA), a hidden reference as high-anchor (HA), and the output of our DeepAFx method with the Inception encoder (Ours). The mid-anchor corresponds to a \textit{tube distortion}~\cite{tubedistref} with drive of +8 dB; a \textit{gate}~\cite{lspgateref} with default settings; and peak normalization to -1dBFS for each task, respectively. All samples for the first two tasks were loudness normalised to -23dBFS. The samples for the music mastering task were monophonic and not loudness normalized so as to evaluate loudness as part of the task. 

The results of the listening test can be seen in \Cref{fig:boxplot} as a violin plot.  For tube amplifier emulation, the median scores in figure order are (0.0, 0.24, 0.96, 0.63, 0.93), respectively. 
The CAFx method ranks above our model, which is reasonable considering it corresponds to a custom, state-of-the-art network. 
Our result for this task is still remarkable, however, since we are only using a multiband compressor in a field where complex models or deep neural proxies have predominated.  
For the non-speech sounds removal task, the median scores in figure order are (0.39, 0.11, 0.89, 0.96), respectively. Our model is rated almost as high as the hidden-anchor, which is again remarkable given the typical expertise and time required for the task. Also note, 1) the mid-anchor is rated lower than the low-anchor because of agitating gating artifacts from Fx processing and 2) the high variance of the low anchor, which contains different speech content (breaths and voice pops) and appears to confuse listeners when rating similarity.
For the automatic music mastering task, the median scores in figure order are (0.15, 0.24, 0.95, 0.88, 0.79). 
Our model is rated above OMS, which we consider state of the art. 
We further conducted multiple post-hoc paired t-tests with Bonferroni correction~\cite{holm1979simple} for each task and condition vs. our method. We found the rank ordering of the results are significant for $p < 0.01$, with the exception of our approach compared to OMS ($p\approx0.0378$).

\section{Conclusion}
\label{sec:conclusion}
We present a new approach to differentiable signal processing that enables the use of stateful, third-party black-box audio effects (plugins) within any deep neural network.
The main advantage of our approach is that it can be used with general, unknown plugins, and can be retrained for a wide variety of tasks by substituting in new audio training data and/or customizing with specific Fx.
We believe that this framework has the potential for broader use, including optimizing legacy signal processing pipelines, style matching, and generative modeling for audio synthesis. 
For future work, we hope to address limitations including optimization of discrete parameters, gradient approx. accuracy, removing the need for paired training data, and robustness to black-box software brittleness.


\pagebreak

\bibliographystyle{IEEEbib}
\bibliography{strings,refs}

\end{document}